\documentclass[11pt]{article}
\usepackage{amssymb}
\usepackage{amsmath}
\usepackage{amsthm}
\usepackage[dvips]{graphics,color}
\usepackage{epsfig}
\usepackage{rotating}
\usepackage{a4wide}
\def\var{\text{Var}}

\def\E{\mathcal{E}}
\def\U{\mathcal{U}}

\begin{document}

\title{Multivariate stochastic volatility modelling using Wishart autoregressive processes}

\author{K. Triantafyllopoulos \\ {\it School of
Mathematics and Statistics, University of Sheffield, Sheffield, S3 7RH, UK}}

\date{\today}

\maketitle

\begin{abstract}

A new multivariate stochastic volatility estimation procedure for financial time series is proposed. A Wishart autoregressive process is considered for the volatility precision covariance matrix, for the estimation of which a two step procedure is adopted. The first step is the conditional inference on the autoregressive parameters and the second step is the unconditional inference, based on a Newton-Raphson iterative algorithm. The proposed methodology, which is mostly Bayesian, is suitable for medium dimensional data and it bridges the gap between closed-form estimation and simulation-based estimation algorithms. An example, consisting of foreign exchange rates data, illustrates the proposed methodology.

\textit{Some key words:} Multivariate volatility, Wishart process, financial time series, covariance, Bayesian forecasting.

\end{abstract}

\section{Introduction}\label{introduction}

Over the last two decades many efforts have been devoted to the development of estimation methods for time-varying volatility and related computational algorithms. Although there is a large literature on univariate volatility estimation methods, it has been widely recognized that multivariate volatility models are required for asset allocation and risk management. Two main classes of models have been identified: (a) multivariate generalized autoregressive conditional heteroscedastic models (GARCH), see e.g. Engle (2002), and (b) multivariate stochastic volatility models (SV), see e.g. Chib {\it et al.} (2006) and Philipov and Glickman (2006). The GARCH family of models deploys maximum likelihood estimation methods, but as it is reported by many authors (see e.g. the review of Bauwens {\it et al.}, 2006) these models suffer from the curse of dimensionality. SV models, which are reviewed in Asai {\it et al.} (2006) and in Yu and Meyer (2006), offer an alternative to the maximum likelihood approach by employing simulation-based Bayesian methods, i.e. Markov chain Monte Carlo (MCMC) or particle filters. However, such estimation proposals may yet under-perform for a number of reasons. Firstly, there are many parameters to consider (perhaps less than in the GARCH specification) and thus the dimensionality problem, which is emphasized for GARCH models, still is an issue. Secondly, the reliance upon simulation-based procedures makes estimation slower and in some occasions more difficult to apply. On this point Brandt and Santa-Clara (2006) state ``While researchers
have explored a variety of numerical solution methods, including solving partial
differential equations, discretizing the state-space, and using Monte Carlo
simulation, these techniques are out of reach for most practitioners and thus
they remain largely in the ivory tower."

The aim of this paper is to develop a multivariate stochastic volatility estimation approach that will bridge the gap between closed-form estimation algorithms, which are found attractive by practitioners, and the sophistication of simulation-based estimation algorithms, which is favoured by many academics. This work contributes fast closed-form estimation procedures, suitable for medium dimensional data, but not compromised on the quality of the estimation considered. The algorithms deployed  in this paper are suitable for real-time application, which more and more is becoming a necessity in financial industry, in particular regarding the implementation of algorithmic trading and related statistical arbitrage strategies (Pole, 2007).

We start in section \ref{model} by considering a Wishart autoregressive stochastic process for the precision of the volatility matrix. Such processes have been introduced in Bru (1991) and further developed as useful probability models for stochastic volatility (Gourieroux, 2006; Gourieroux {\it et al.}, 2009). In this paper we develop an extension of Uhlig (1994) matrix variate random walk model in order to develop inference conditionally on the autoregressive (AR) parameters of the Wishart autoregressive process. Under this framework, we show that the volatility process is also autoregressive and we determine its parameters as functions of the parameters of the AR precision process. Assuming that the AR parameters of the precision process are stochastic, we identify their posterior distribution (up to a proportionality constant) and we propose approximating its mode by using a Newton-Raphson iterative procedure. Thus we arrive to estimating the volatility covariance matrix by conjugate Bayesian methods and the AR parameters of the Wishart process by iterative methods. Section \ref{diagnostics} discusses three diagnostic criteria, namely the log posterior function, Bayes factors and minimum time-averaged portfolio risk. By considering an AR process for the precision of the volatility, this paper aims to overcome the limitation of random walk evolution proposed in Uhlig (1994) and adopted in a number of studies (Quintana {\it et al.}, 2003; Soyer and Tanyeri, 2006; Triantafyllopoulos, 2008).

The proposed methodology is illustrated by Monte Carlo simulations as well as by data consisting of foreign exchange rates (FX) of five currencies \emph{vis-$\grave{a}$-vis} the US dollar. Our empirical results suggest that the proposed volatility estimators have low computational cost, considering similar computational algorithms, such as those in Philipov and Glickman (2006) and in relevant studies mentioned below. The dimensionality of the FX data is in par with similar recent studies in the literature, e.g. Dan\'{i}elsson (1998) considers 4-dimensional data, Liesenfeld and Richard (2003) consider 4-dimensional data, Philipov and Glickman (2006) consider 5-dimensional data, and Chib {\it et al.} (2006) consider 10-dimensional data. In our empirical study we find that the proposed methodology compares well with the random walk variance models of Soyer and Tanyeri (2006) (similar models have been presented in Quintana and West (1987), Quintana {\it et al.} (2003), Triantafyllopoulos (2008)) and with the dynamic conditional correlation GARCH models of Engle (2002). Finally, the paper concludes in section \ref{conclusions} with closing comments.

\section{Description of the model}\label{model}

Consider the $p$-dimensional time series vector $\{y_t\}$, consisting typically of log returns or arithmetic returns of prices of assets or foreign exchange rates or of any other relevant financial instrument. For example, if $p_t=(p_{1t},\ldots,p_{pt})'$ denotes the $p$-dimensional column vector of prices of a list of $p$ assets or the value of $p$ exchange rates at time $t$, the log returns are defined as $y_{it}=\log p_{it}-\log p_{i,t-1}$ and the  arithmetic returns are defined as $y_{it}=p_{it}/p_{i,t-1}-1$, for $y_t=(y_{1t},\ldots,y_{pt})'$ and $t\geq 2$. A classical modelling setting for $\{y_t\}$ is to assume that conditionally on a volatility matrix $\Sigma_t$, which is the main theme of econometric analysis and subject to estimation, the distribution of $y_t$ is multivariate normal, i.e.
\begin{equation}\label{model1}
y_t=\mu+\Sigma_t^{1/2} \epsilon_t , \quad \epsilon_t \sim N_p(0, I_p),
\end{equation}
where $\mu$ denotes a historical mean vector, $\Sigma_t^{1/2}$ denotes the square root matrix of $\Sigma_t$, and the sequence of $\{\epsilon_t\}$ follows a $p$-dimensional Gaussian white noise process with unit diagonal variances (here $I_p$ denotes the $p\times p$ identity matrix).

In order to define a stochastic evolution for $\{\Sigma_t\}$, first we assume that for all $t$, the $p\times p$ precision covariance matrix $\Phi_t=\Sigma_t^{-1}$ exist, i.e. $\Sigma_t$ is strictly positive definite, and subsequently it is assumed that $\{\Phi_t\}$ follows Uhlig's Wishart autoregressive process of order one
(Uhlig, 1994, 1997),
\begin{equation}\label{model3}
\Phi_t = k A \U (\Phi_{t-1})'B_t\U(\Phi_{t-1})A'+\Lambda_t,
\end{equation}
where $k$ is a constant to be determined, $A$ is a $p\times p$ autoregressive parameter matrix, $\Lambda_t$ is a $p\times p$ symmetric matrix and $\U(\Phi_{t-1})$ denotes the upper triangular matrix of the Choleski decomposition of the matrix $\Phi_{t-1}$. In most practical applications, $\Lambda_t=0$, as it is used in section \ref{empiricalresults} of this paper, but as it is shown in page \pageref{model1:psi1} below, $\Lambda_t\neq 0$ need to be considered to accommodate for Wishart AR processes of higher order than one. In the above model formulation, the $p\times p$ matrix $B_t$ follows, independently of $\Phi_{t-1}$, a singular multivariate beta distribution with parameters $a/2$ and $b/2$, written $B_t\sim B_p(a/2,b/2)$; below and in the next section we discuss about the parameters $a,b$.

To motivate model \eqref{model3}, suppose $A=I_p$ and $\Lambda_t=0$, so that \eqref{model3} is reduced to the random walk evolution considered in Uhlig (1994), i.e. $\Phi_t=\Phi_{t-1}+\E_t$, where $\E_t$ is a symmetric random matrix with expectation zero, which supports the random walk property $E(\Phi_t|\Phi_{t-1})=\Phi_{t-1}$. In the appendix we discuss in some detail Uhlig's random walk model, as well as the singular beta distribution. The parameters $a,b,k$, which all are set to take particular values (see below), depend on a forgetting or discount factor $0<\delta<1$, which controls the magnitude of the shocks introduced as we move from $\Phi_{t-1}$ to $\Phi_t$, so that the only free parameter is $\delta$ (the specification of $a,b,k$ is discussed in detail in section \ref{inference:onA}). The parameters $a$ an $b$ are conveniently chosen ($a$ is a function of $\delta$ and $b=1$) so that $E(B_t)=k^{-1}I_p$, in order to support the random walk property $E(\Phi_t|\Phi_{t-1})=\Phi_{t-1}$; Uhlig (1994) shows that $B_t$ has to follow a singular beta distribution for this to be possible, in order to have $b<p-1$ (because for a non-singular distribution $b$ is greater than $p-1$). Furthermore, we are happy to allow for $b<p-1$, because a given medium dimension $p$ is assumed.

Evolution \eqref{model3} has some similarities with the random walk models considered in Soyer and Tanyeri (2006) and Triantafyllopoulos (2008). These authors use a different model for the mean process $\mu$ (Soyer and Tanyeri (2006) use exponential smoothing and Triantafyllopoulos (2008) uses a state space model) and instead of $k$ in \eqref{model3} they use $1/\delta$. With this setting it is claimed that in their random walk process the expectations are preserved, i.e. the prior expectation of $\Phi_t$ at $t$ equals to the posterior expectation of $\Phi_{t-1}$ at $t-1$. In this paper, in section \ref{inference:onA}, we show that this is incorrect and that choice basically results in a shrinkage-type evolution for $\{\Phi_t\}$, which is unrealistic. We thus show that one needs to consider a particular expression of $k$, essentially given as a function of $\delta$, but different to $1/\delta$. Triantafyllopoulos (2008) extends the approach of Soyer and Tanyeri (2006), to include several discount factors. This approach suffers too from the above mentioned deficiency and with more discount factors introduced in the model, their estimation or specification may slow down the speed of the algorithm. Compared to the above studies, this paper suggests an autoregressive evolution for $\{\Phi_t\}$.

Considering model \eqref{model3}, we can see that $E(\Phi_t|\Phi_{t-1})=A\Phi_{t-1}A'+\Lambda_t$, since from the beta distribution it is $E(B_t)=k^{-1}I_p$. From this property, and with analogy to the random walk model described above, we can write $\Phi_t=A \Phi_{t-1} A'+\E_t$, where $\{\E_t\}$ is a sequence of symmetric random matrices with expectation $E(\E_t)=\Lambda_t$. Both models \eqref{model3} and \eqref{model2} produce the same $E(\Phi_t|\Phi_{t-1})$; model \eqref{model3} uses a multiplicative law, while \eqref{model2} uses an additive law. In fact one may consider a higher order AR model, defined by
\begin{equation}\label{model2}
\Phi_t=\sum_{j=1}^d A_j \Phi_{t-j} A_j' + \E_t, \quad
t=d,d+1,\ldots,N,
\end{equation}
where $A_1,\ldots,A_d$ are $p\times p$ parameter matrices and $d$ is the autoregression order. We call \eqref{model2} as Uhlig's Wishart autoregressive process (UWAR$(d)$) because it can be written as a UWAR(1) process (see the next section) and we adopt Uhlig's multiplicative evolution \eqref{model3} for inference. Process \eqref{model2} should not be confused with the Wishart autoregressive processes, proposed by Gourieroux {\it et al.} (2009), in which the sequence $\{\E_t\}$ is i.i.d, while in \eqref{model2} it can be shown that $\{\E_t\}$ is conditionally heteroscedastic (Soyer and Tanyeri, 2006, p. 982).

The volatility model is thus defined by the observation equation \eqref{model1} and the evolution of the process $\{\Phi_t\}$ \eqref{model3}. Finally, it is assumed that initially, $\Phi_0$ follows a Wishart distribution with some known degrees of freedom $n_0>p-1$ and scale matrix $F_0$, written as $\Phi_0\sim W_p(n_0,F_0)$.

Let $D_t=(y_1,\ldots,y_t)$ denote the data or information set at time $t$, comprising observed data vectors $y_1,\ldots,y_t$, for $t=1,\ldots,N$. We wish to obtain the posterior distribution of $\Phi_t$, given $D_t$. The model parameters are $A$ (the AR parameter matrix), $\Lambda_t$ (the mean of $\E_t$), and $\delta$ the discount factor. A fully Bayesian approach would require the specification of the priors $\Phi_1,\ldots,\Phi_t$, given $A$, $\Lambda_i$ and $\delta$ and the priors of $A$, $\Lambda_i$ and $\delta$ and it should rely on MCMC. Such an approach has been proposed by Philipov and Glickman (2006), who use a Gibbs sampler to sample from the posterior of $\Phi_t$, while the hyperparameters of their model are estimated by a Metropolis-Hastings algorithm.

In this paper, as our aim is to bridge the gap between closed-form estimation and simulation-based estimation algorithms, we adopt a \emph{two step} estimation procedure. In the first step, conditionally on $A,\Lambda_t,\delta$, we obtain the posterior distribution of $\Phi_t$, and in the second step we obtain the posterior distribution of $A$. Then, in order to obtain a working estimator of $A$, we resort to a Newton-Raphson method to approximate the mode of the posterior distribution of $A$. $\Lambda_t$ is assumed known, here it is set to the zero matrix, which is suitable for the AR representation supporting the expectation $E(\Phi_t|\Phi_{t-1})=A\Phi_{t-1}A'$. We note that in theory $\Lambda_t=0$, could cause $\Phi_t$ in \eqref{model2} to be too close to the zero matrix, but in application we have found this is not a problem as at each time $t$, $A$ balances this effect. For a higher autoregressive order $d>1$, $\Lambda_t=E(\E_t)$ is a non-zero mean, as it is evident from equation \eqref{model1:psi1} in section \ref{inference:onA} below. In line with other authors, for the specification of the discount factor $\delta$, we adopt a non-Bayesian setting. $\delta$ is responsible for the magnitude of the shocks in $\Phi_t$, incurred from $t-1$ to $t$. At the one end, $\delta=1$ implies $\Phi_t=A\Phi_{t-1}A'$, or $\E_t=0$ or $B_t=I_p$ (with probability 1), and at the other end a low value of $\delta$  introduces large shocks to the $\Phi_t$ process. Quintana and West (1987) and Soyer and Tanyeri (2006), considering random walk models, suggest values of $\delta$ around 0.8 or 0.9. Below we show that $\delta$ must satisfy $2/3<\delta<1$, for the volatility process to make sense. With the above setting in place, the posterior distribution of $\Phi_t$ has been implicitly conditioned on the mode of $A$ and on a given particular value of $\delta$. We assume a matrix-variate normal prior distribution for $A$, i.e. $A\sim N_{p\times p}(M_A,V_A,W_A)$, where $M_A$ is a $p\times p$ matrix mean, $V_A$ is a $p\times p$ left covariance matrix and $W_A$ a $p\times p$ right covariance matrix. This means that $\textrm{vec}(A)$ follows a $p^2$-dimensional Gaussian distribution, or $\textrm{vec}(A)\sim N_{p^2}(\textrm{vec}(M_A),W_A\otimes V_A)$, where $\otimes$ denotes the Kronecker operator.

\section{Inference}\label{inference}

\subsection{Inference conditional on $A$}\label{inference:onA}

\subsubsection{Case of AR order $d=1$}

First we discuss inference for AR order $d=1$. The derivation of the posterior distribution of $\Phi_t$ is inductive. Conditionally on $A$, assume that $\Phi_{t-1}$ has the posterior distribution $\Phi_{t-1}|A,D_{t-1}\sim W_p(n+p-1,F_{t-1})$, where $F_{t-1}$ implicitly depends on $A$ and $n=(1-\delta)^{-1}$, for a discount or forgetting factor $0<\delta<1$. Starting at $t=1$, this is consistent with the prior of $\Phi_0$, if we set $n_0=n+p-1$. In order to set up the prior and posterior distributions of $\Phi_t$ and to calculate the value of $k$ (see equation \eqref{model3}), we first consider the case of $\Lambda_t=0$. If we then specify $a=\delta (1-\delta)^{-1}+p-1$ and $b=1$, we see from Uhlig (1994) that $k^{-1}A^{-1}\Phi_t|A,D_{t-1}\sim W_p(\delta n+p-1,F_{t-1})$, or $\Phi_t|A,D_{t-1}\sim W_p(\delta n+p-1,kAF_{t-1}A')$; details of this argument are discussed in the appendix. From the above it is $E(\Phi_{t-1}|A,D_{t-1})=(n+p-1)F_{t-1}$
and
$E(\Phi_t|A,D_{t-1})=(\delta n+p-1)kAF_{t-1}A'$, and so by
equalizing these two expectations we obtain
$$
k=\frac{n+p-1}{\delta n+p-1}=\frac{\delta(1-p)+p}{\delta(2-p)+p-1}.
$$
Under the above setting, this value of $k$ guarantees the autoregressive property of the model, expressed by $E(\Phi_t|A,D_{t-1})=AE(\Phi_{t-1}|A,D_{t-1})A'$.

We note that, considering the random walk model $(A=I_p)$, West and Harrison (1997, Chapter 16) and Soyer and Tanyeri (2006) use $k=1/\delta$. Although it is easily verified that this is a
correct choice for $p=1$, setting $k=1/\delta$ for $p>1$ results
in a shrinkage-type evolution for $\{\Phi_t\}$. This can be
seen by first noting that, with $k=1/\delta$, we have $E(\Phi_t|D_{t-1})-E(\Phi_{t-1}|D_{t-1})=(p-1)(\delta^{-1}-1)F_{t-1}$
and therefore, the expectation is not preserved from time $t-1$ to
$t$, as we have
$E(\Phi_t|D_{t-1})>E(\Phi_{t-1}|D_{t-1})$. In particular, when $p$ is large, even if $\delta\approx 1$, the
above model postulates that the estimate of $\Phi_t$ is
larger than that of $\Phi_{t-1}$;
such a setting is clearly inappropriate. Triantafyllopoulos (2008) proposes the use of $p$ discount factors $\delta_1,\ldots,\delta_p$ to replace the single value of $\delta$, but this choice too results in $E(\Phi_t|D_{t-1})>E(\Phi_{t-1}|D_{t-1})$, which is not in agreement with the claimed random walk evolution of $\Phi_t$. In this paper we suggest to use a single forgetting factor $\delta$ because (a) this enables the definition of $k$ as above, in order to preserve the expectations in the random walk model and (b) the use of $p$ discount factors may introduce estimation difficulties, because $p$ discount factors would need to be estimated or specified.

We note that $a>p-1$, but $1=b<p-1$, the latter of which being responsible for the singularity of the beta distribution. The singular beta density, being defined on the Stiefel manifold, replaces the determinant of $I_p-B_t$ (which is zero) by the only positive eigenvalue of that matrix (due to $b=1$). On the other hand, the determinant of $B_t$ remains positive as $a>p-1$ and thus all $p$ eigenvalues of $B_t$ are positive; this beta distribution is briefly discussed in the appendix. In the general case of $\Lambda_t\neq 0$, the prior of $\Phi_t$ becomes $\Phi_t|A,D_{t-1}\sim W_p(\delta n+p-1,kAF_{t-1}A'+\Lambda_t)$.

So far our discussion has been focused on the precision process $\{\Phi_t\}$. Before we proceed with inference, we show that the volatility $\{\Sigma_t\}$ follows an autoregressive process too. Without loss in generality and for convenience in the exposition, we assume $\Lambda_t=0$; this setting is appropriate for $d=1$ and for $d>1$ the amendments are minor. From \eqref{model3} we have
$$
\mathcal{E}_t=\Phi_t-A\Phi_{t-1}A'=kA\U(\Phi_{t-1})'B_t\U(\Phi_{t-1})A'-A\Phi_{t-1}A'.
$$
Applying the matrix inversion lemma in \eqref{model2} we have
\begin{equation}\label{sigma:ar1}
\Sigma_t=\Phi_t^{-1}=(A\Phi_{t-1}A'+\mathcal{E}_t)^{-1}=(A')^{-1}\Sigma_{t-1}A^{-1} Y,
\end{equation}
where using \eqref{model3}, $Y=(\mathcal{E}_t(A')^{-1}\Sigma_{t-1}A^{-1}+I_p)^{-1}=k^{-1}A\Phi_{t-1}(\U(\Phi_{t-1}))^{-1}B_t^{-1}(\U(\Phi_{t-1})')^{-1}A^{-1}$. Thus
\begin{equation}\label{ex:Y}
E(Y|\Sigma_{t-1})=k^{-1}A\U(\Phi_{t-1})'E(B_t^{-1})(\U(\Phi_{t-1})')^{-1}A^{-1}=
\frac{\delta(1-\delta)^{-1}-1}{k(\delta(1-\delta)^{-1}-2)}I_p=cI_p.
\end{equation}
This result is established by noting that with the stated beta distribution of $B_t$, $B_t^{-1}-I_p$ follows a type II singular multivariate beta distribution (D\'{i}az-Garc\'{i}a and Guti\'{e}rrez, 2008). From this we obtain $E(B_t^{-1}-I_p)=b(a-p-1)^{-1}I_p$ and $E(B_t^{-1})=(a+b-p-1)(a-p-1)^{-1}I_p$, with $a=\delta(1-\delta)^{-1}+p-1$ and $b=1$. For more details on the derivations of moments of the type II beta distribution see Khatri and Pillai (1965) and Konno (1988). The above expectation is valid only for $a>p+1$, or $\delta>2/3$, which will be assumed henceforth in this paper. Therefore, given $\Sigma_{t-1}$, and combing \eqref{sigma:ar1} and \eqref{ex:Y}, we obtain $E(\Sigma_t|\Sigma_{t-1})=c(A')^{-1}\Sigma_{t-1}A^{-1}$ and thus by defining $C=c^{1/2}(A')^{-1}$, $\{\Sigma_t\}$ follows an AR process, i.e.
\begin{equation}\label{volprocess:sigma}
\Sigma_t=C\Sigma_{t-1}C'+ \mathcal{Z}_t,
\end{equation}
for some symmetric random matrix $\mathcal{Z}_t$ with zero mean matrix.

Having established the prior $\Phi_t|A,D_{t-1}\sim W_p(\delta n+p-1,kAF_{t-1}A'+\Lambda_t)$, the posterior distribution follows by a similar argument as in Triantafyllopoulos (2008)
\begin{equation}\label{postPhi1}
\Phi_t|A,D_t\sim W_p(n+p-1,F_t),
\end{equation}
where $e_t=y_t-\mu$ is the residual vector and $F_t=(e_te_t'+(kAF_{t-1}A'+\Lambda_t))^{-1}$. From the above reference, the one-step forecast distribution of $y_t$, is a $p$-variate Student $t$ distribution with $\delta n$ degrees of freedom and spread matrix $\delta^{-1}n^{-1}(kAF_{t-1}A'+\Lambda_t)^{-1}$, i.e. $y_t|A,D_{t-1}\sim t_p(\delta n, \mu, \delta^{-1}n^{-1}(kAF_{t-1}A'+\Lambda_t)^{-1})$.

\subsubsection{Case of AR order $d\geq 1$}

The above results assume first order UWAR processes, i.e. $d=1$. Consider now the general case of $d\geq 1$. From the autoregression (\ref{model2}) it is easy to verify
\begin{gather*}
\left[ \begin{array}{cccc} \Phi_t & 0 & \cdots & 0 \\ 0 & \Phi_{t-1} & \cdots & 0 \\ \vdots & \vdots & \ddots & \vdots \\ 0 & 0 & \cdots & \Phi_{t-d+1} \end{array}\right]    =   \left[ \begin{array}{ccccc} A_1 & A_2 & \cdots & A_{d-1} & A_d \\ I_p & 0 & \cdots & 0 & 0 \\ \vdots & \vdots & \ddots & \vdots & \vdots \\ 0 & 0 & \cdots & I_p & 0 \end{array}\right] \left[ \begin{array}{cccc} \Phi_{t-1} & 0 & \cdots & 0 \\ 0 & \Phi_{t-2} & \cdots & 0 \\ \vdots & \vdots & \ddots & \vdots \\ 0 & 0 & \cdots & \Phi_{t-d} \end{array}\right] \\ \times \left[\begin{array}{ccccc} A_1' & I_p & 0 & \cdots & 0 \\ A_2' & 0 & I_p & \cdots & 0 \\ \vdots & \vdots & \vdots & \ddots & \vdots \\ A_d' & 0 & 0 & \cdots & 0 \end{array}\right] + \left[\begin{array}{cccc} \E_t & -A_1\Phi_{t-1} & \cdots & -A_{d-1}\Phi_{t-d+1} \\ -\Phi_{t-1}A_1' & 0 & \cdots & 0 \\ \vdots & \vdots & \ddots & \vdots \\ -\Phi_{t-d+1}A_{d-1}' & 0 & \cdots & 0 \end{array}\right],
\end{gather*}
which can be written as
\begin{equation}\label{model1:psi1}
\Psi_t=A\Psi_{t-1}A'+E_t.
\end{equation}

Furthermore, from the identity
$$
\Phi_t=[I_p,0,\ldots,0] \left[ \begin{array}{cccc} \Phi_t & 0 & \cdots & 0 \\ 0 & \Phi_{t-1} & \cdots & 0 \\ \vdots & \vdots & \ddots & \vdots \\ 0 & 0 & \cdots & \Phi_{t-d+1} \end{array}\right] \left[\begin{array}{c} I_p \\ 0 \\ \vdots \\ 0\end{array}\right],
$$
we can write $\Phi_t=J\Psi_t J'$, where $J=[I_p,0,\ldots,0]$ and also we can verify that $\Sigma_t^{1/2}=J\Psi_t^{-1/2}J'$. Thus equation \eqref{model1} can be written as
\begin{equation}\label{model1:psi2}
y_t=\mu+J\Psi_t^{-1/2}J'\epsilon_t.
\end{equation}
By assuming that $\Psi_0$ follows a Wishart distribution, the posterior distribution of $\Psi_t|A,D_t$ is a Wishart and from the block diagonal construction of $\Psi_t$ we have that $\Phi_t|A,D_t$ will follow a Wishart distribution too. Then the one-step ahead forecast distribution of $y_t$ is a Student $t$. These results are conditional on $A_1,\ldots,A_d$ or conditional on $A$. Inference unconditional on $A$ is obtained if we make use of the above transformation and work with $d=1$, which is developed next.

\subsection{Inference unconditional on $A$}\label{inference:notA}

Let $A$ be a $p\times p$ non-singular stochastic matrix. From the joint prior density $f(\Phi_t,A|D_{t-1})=f(\Phi_t|A,D_{t-1})f(A|D_{t-1})$ and from an application of Bayes theorem for $(\Phi_t,A)$, we have
$f(\Phi_t,A|D_t)\propto f(y_t|\Phi_t) f(\Phi_t|A,D_{t-1})f(A|D_{t-1})$, so that
\begin{equation}\label{int1}
f(A|D_t)\propto f(A|D_{t-1}) \int f(y_t|\Phi_t)f(\Phi_t|A,D_{t-1})\,d\Phi_t.
\end{equation}
From the forecast distribution of $y_t$, the integral of \eqref{int1} is
$$
\int f(y_t|\Phi_t)f(\Phi_t|A,D_{t-1})\,d\Phi_t \propto |e_te_t'+(kAF_{t-1}A'+\Lambda_t)^{-1}|^{-(\delta n+p)/2},
$$
and so
$$
f(A|D_t)\propto f(A) \prod_{j=1}^t |e_je_j'+(kAF_{j-1}A'+\Lambda_j)^{-1}|^{-(\delta n+p)/2},
$$
where $f(A)$ is the prior density of $A$.

In order to find the mode $\hat{A}$ of $f(A|D_t)$, we note that the matrix equation $\partial f(A|D_t)/\partial A=0$ (with respected to $A$; here $\partial f(\cdot)/\partial$ denotes first partial derivative) does not appear to admit an analytical solution. Thus, we approximate the true mode $\hat{A}$, by employing the Newton-Raphson method, according to which at each time $t$, for iteration $i=1,2,\ldots$, we compute $\hat{A}^{(i)}$ using the formula
\begin{equation}\label{NR1}
\textrm{vec}(\hat{A}^{(i)}) = \textrm{vec}(\hat{A}^{(i-1)}) + \left( \frac{\partial^2 \log f(A|D_t) }{ \partial \textrm{vec}(A) \partial \textrm{vec}(A)' } \right)^{-1}\bigg|_{A=\hat{A}^{(i-1)}} \frac{\partial \log f(A|D_t)}{\partial \textrm{vec}(A)} \bigg|_{A=\hat{A}^{(i-1)}},
\end{equation}
where $\hat{A}^{(0)}$ is initially given and $\textrm{vec}(\cdot)$ denotes the column stacking operator of an unrestricted matrix. Under some regulatory conditions (Sumway and Stoffer, 2006, \S6.3), the algorithm converges to the true mode $\hat{A}$.

The density of $\log f(A|D_t)$ is
$$
\log f(A|D_t) = \log c + \log f(A) -\frac{\delta n+p}{2}\sum_{j=1}^t\log |e_je_j'+(kAF_{j-1}A'+\Lambda_j)^{-1}|,
$$
where $c$ is the proportionality constant of $f(A|D_t)$. Then, the first partial derivative of $\log f(A|D_t)$ with respect to $A$ is
\begin{eqnarray}
\frac{\partial\log f(A|D_t)}{\partial A} &=& \frac{\partial\log f(A)}{\partial A} - k(\delta n+p)\sum_{j=1}^t ( e_je_j'((kAF_{j-1}A'e_je_j'\nonumber \\ &&+\Lambda_je_je_j'+I_p)^{-1})-
(kAF_{j-1}A'+\Lambda_j)^{-1})AF_{j-1}.\label{mode1}
\end{eqnarray}
In the appendix it is shown that for an unrestricted matrix of variables $X$ and for constant symmetric matrices $B$, $C$ and $G$, it is
\begin{equation}\label{deriv}
\frac{\partial \log |BXCX'+BG+I_p|}{\partial X} = 2B(XCX'B+BG+I_p)^{-1}XC,
\end{equation}
so that
\begin{eqnarray*}
\frac{\partial \log |B+(XCX'+ G)^{-1}| }{\partial X} &=& \frac{\partial \log |(XCX'+G)^{-1}(I_p+BXCX'+BG)|}{\partial X}\\ &=&  \frac{\partial \log |BXCX'+BG+I_p|}{\partial X} -
\frac{\partial \log |XCX'+G|}{\partial X} \\ &=& 2(B(XCX'B+GB+I_p)^{-1}-(XCX'+G)^{-1})XC,
\end{eqnarray*}
which by substituting $X=A$, $B=e_je_j'$, $C=kF_{j-1}$, and $G=\Lambda_j$, immediately gives the expression for the derivative of $\log f(A|D_t)$. Expression \eqref{deriv} extends previous results on the partial derivative of the logarithm of the determinant of a symmetric matrix (Harville, 1997, p. 327).

From the prior density of $A$ we have
\begin{equation}\label{mode3}
\frac{\partial \log f(A)}{\partial A} = -V_A^{-1}(A-M_A)W_A^{-1},
\end{equation}
so that derivative \eqref{mode1} becomes
\begin{eqnarray}
\frac{\partial\log f(A|D_t)}{\partial A} &=& -V_A^{-1}(A-M_A)W_A^{-1} - k(\delta n+p)\sum_{j=1}^t ( e_je_j'((kAF_{j-1}A'e_je_j'\nonumber \\ &&+\Lambda_je_je_j'+I_p)^{-1})-
(kAF_{j-1}A'+\Lambda_j)^{-1})AF_{j-1},\label{mode4}
\end{eqnarray}
which, by applying the $\textrm{vec}(\cdot)$ operator, gives the gradient in the right hand side of \eqref{NR1}, i.e.
\begin{eqnarray}
\frac{\partial \log f(A|D_t)}{\partial \textrm{vec}(A)} &=& -(W_A^{-1} \otimes V_A^{-1} ) (\textrm{vec}(A)-\textrm{vec}(M_A)) \nonumber \\ && -k(\delta n+p) \sum_{j=1}^t \bigg( (F_{j-1}\otimes e_je_j') \textrm{vec}(kAF_{j-1}A'e_je_j'+\Lambda_je_je_j'+I_p)^{-1})A \nonumber \\ &&  - (F_{j-1}\otimes I_p) \textrm{vec}(kAF_{j-1}A'+\Lambda_j)^{-1}A \bigg). \label{NR2}
\end{eqnarray}
To obtain the Hessian matrix of \eqref{NR1} we differentiate \eqref{NR2}, i.e.
\begin{gather}
\frac{\partial^2 \log f(A|D_t) }{ \partial \textrm{vec}(A) \partial \textrm{vec}(A)' } = -W_A^{-1}\otimes V_A^{-1} +k(\delta n +p) \sum_{j=1}^ t \bigg( (F_{j-1}\otimes e_je_j') \nonumber \\  \times (kF_{j-1}A'e_je_j'+A^{-1}\Lambda_je_je_j'+A^{-1})^{-1}   \otimes (kF_{j-1}A'e_je_j'+A^{-1}\Lambda_je_je_j'+A^{-1})^{-1} \nonumber \\  \times ((e_je_j'\otimes kF_{j-1})K_p-e_je_j'\Lambda_j A^{-1}\otimes A^{-1} - A^{-1} \otimes A^{-1})   -(F_{j-1}\otimes I_p) (kF_{j-1}A'+A^{-1}\Lambda_j)^{-1} \nonumber \\ \otimes (kF_{j-1}A'+A^{-1}\Lambda_j)^{-1}  ((I_p\otimes kF_{j-1})K_p-\Lambda_jA^{-1}\otimes A^{-1}) \bigg), \label{NR3}
\end{gather}
where $K_p$ is the $p^2\times p^2$ vec-permutation matrix, i.e. $\textrm{vec}(A')=K_p\textrm{vec}(A)$.

This result follows from standard matrix differentiation rules, e.g. for $X$ being a matrix of unrestricted variables and $F(X)$ a non-singular matrix of functions of $X$, it is
$$
\frac{\partial \textrm{vec}( F(X)^{-1}) }{ \partial \textrm{vec}(X)} = -F(X)^{-1}\otimes F(X)^{-1} \frac{\partial \textrm{vec}(F(X)) }{ \partial \textrm{vec}(X)},
$$
for a proof of which the reader is referred to Harville (1997, \S16.6). With \eqref{NR2} and \eqref{NR3} in place, at each iteration $i=1,2,\ldots$, we can compute $\hat{A}^{(i)}$ from \eqref{NR1}. Initially we set $A^{(0)}=I_p$, although, in our experience this is not critical for convergence. Convergence is assumed at iteration $i$, for which $\parallel A^{(i)}-A^{(i-1)}\parallel_2\leq Tol$, for some small tolerance value $Tol$, where $\parallel \cdot \parallel_2$ denotes the Frobenius norm; similar stoppage rules are discussed in Shumway and Stoffer (2006, \S6.3). Note that typically not many iterations are needed for convergence, although this may depend on the specific application and on the dimension of the data. Also, note, that since $f(A|D_t)$ is a symmetric distribution, the computed approximation $\hat{A}$ provides an approximation of the mean matrix $E(A|D_t)$ too.

The posterior distribution of $\Phi_t$ is given by
\begin{gather*}
f(\Phi_t|D_t) = \int f(\Phi_t|A,D_t)f(A|D_t)\,dA \\ \propto  |\Phi_t|^{(n-2)/2}  \int \exp(\textrm{trace}(-F_t^{-1}\Phi_t/2)) \prod_{j=1}^t |e_te_t'+(kAF_{j-1}A'+\Lambda_j)^{-1}|^{-(\delta n+p)/2} f(A) \,dA.
\end{gather*}
The above integral is not easy to calculate in closed form, but one option is to apply simulation-based or numerical methods for its evaluation. Another option, which is deployed in section \ref{empiricalresults}, is to use the Wishart posterior $\Phi_t|A=\hat{A},D_t\sim W_p(n+p-1, \hat{F}_t)$, where $\hat{F}_t$ is the estimated value of $F_t$ if we replace $A$ by $\hat{A}$. Similarly, we can work with the prior distribution of $\Phi_t|A=\hat{A},D_{t-1}$ and the forecast distribution of $y_t|A=\hat{A},D_{t-1}$, where now the computation of $\hat{A}$ uses data up to time $t-1$ or information $D_{t-1}$.

\section{Diagnostics}\label{diagnostics}

Diagnostic tools comprise Bayesian and non-Bayesian. For example, from a Bayesian perspective Bayes factors, Schwartz's criterion (also known as Bayesian information criterion), Bayesian deviance and model averaging are all available within a model choice framework. From a classical perspective, the likelihood function and criteria such as mean absolute deviation and mean square error are also available. Bayesian model choice criteria, such as those mentioned above, are covered in detail in Robert (2007, Chapter 7). The advantage of the Bayesian approach is its capability of taking into account not only the data, but also prior information. However, some of the above criteria involve the use of simulation-based methods, such as deviance and model averaging. Schwartz's criterion uses a Laplace approximation of the Bayes factor, but this criterion is not relevant for comparison of models having the same number of parameters or of models that are not nested one to other. The issue of incorporating prior information is not so critical, since prior information in time series has the tendency to deflate over time. In this paper, as we propose a methodology to bridge the gap between closed-form estimation and simulation-based algorithms, we do not discuss model choice criteria that rely upon simulation. Next, we discuss three model comparison criteria, namely the log-posterior, Bayes factors and minimum time-averaged portfolio risk. These three criteria aim at comparing models of the same form of model \eqref{model3} for different model components, such as discount factors.

\subsection{Log-posterior function}

The log likelihood function can be obtained by using the classical error decomposition for state space models, i.e. based on information $D_N=(y_1,\ldots,y_N)$, the likelihood is $L=\prod_{t=1}^N f(y_t|A,D_{t-1})$, which is a product of $N$ Student $t$ densities. However, since the focus in this paper is on the estimation of $\Sigma_t$ and in $L$ this is only indirectly involved, in the sequel we discuss the log posterior function instead.

Based on information $D_N$, the log posterior function (Fahrmeir, 1992) of the volatilities $\Sigma_1,\ldots,\Sigma_N$, may be used as a means of model comparison as well as it can be used to choose the hyperparameter $\delta$. Write $\Sigma_N^*=(\Sigma_1,\ldots,\Sigma_N)$, then, by using Bayes theorem, the posterior of $\Sigma_N^*$ is
\begin{eqnarray}
f(\Sigma_N^*|A,D_N) &=& f(y_N|\Sigma_N)f(\Sigma_N^*|A,D_{N-1})  = c_N^N f(\Sigma_{N-1}^*|A,D_{t-1}) f(y_N|\Sigma_N) f(\Sigma_N|\Sigma_{N-1},A) \nonumber \\ &=& c_1^N f(\Sigma_0|A)  \prod_{t=1}^N f(y_t|\Sigma_t)f(\Sigma_t|\Sigma_{t-1},A), \label{logp1}
\end{eqnarray}
where $c_1^N=\prod_{t=1}^N (f(y_t|D_{t-1},A))^{-1}$. Since $c_1^N$ does not depend on $\{\Sigma_t\}$, we exclude it from the computation of the posterior, i.e. we set $c_1^N=1$, but if we wish to estimate $A$ using the principle of log posterior maximization, then $c_1^N$ has to be included as it implicitly depends on $A$. From \eqref{model1} we have $y_t|\Sigma_t\sim N_p(\mu,\Sigma_t)$. Below we derive the density $f(\Sigma_t|\Sigma_{t-1},A)$.

First we derive the density $f(\Phi_t|\Phi_{t-1},A)$. From \eqref{model3} we have $B_t=k^{-1}\U(\Phi_{t-1})'^{-1}A^{-1}(\Phi_t-\Lambda_t)A'^{-1}\U(\Phi_{t-1})'$, from which and D\'{i}az-Garc\'{i}a and Guti\'{e}rrez (1997, Theorem 1) the Jacobian of $B_t$ with respect to $\Phi_t$ is
$$
(\,dB_t)=|B_t|^{p/2}|\Phi_t-\Lambda_t|^{p/2}|k^{-1}\U(\Phi_{t-1})'^{-1}A^{-1}|(\,d\Phi_t).
$$
Thus, from the stated beta distribution $B_t\sim B_p(a/2,1/2)$, with density
$$
f(B_t)\pi^{(1-p)/2} \frac{\Gamma_p((a+1)/2)}{\Gamma(1/2)\Gamma_p(a/2)} \xi_t^{-p/2} |B_t|^{(a-p-1)/2},
$$
for $a=\delta(1-\delta)^{-1}+p-1$ (see section \ref{inference:onA}),
the density of $\Phi_t|\Phi_{t-1},A$ is
$$
f(\Phi_t|\Phi_{t-1},A)=\pi^{-p/2} \frac{\Gamma_p((a+1)/2)}{\Gamma(1/2)\Gamma_p(a/2)} \xi_t^{-p/2} k^{-3p} |\Sigma_{t-1}|^{(p+3)/2} |A|^{-(p+4)} |\Sigma_t^{-1}-\Lambda_t|^{p+1},
$$
where $\xi_t$ is the only positive eigenvalue of $I_p-B_t$.

Since $\Sigma_t=\Phi_t^{-1}$, and the Jacobian of $\Phi_t$ with respect to $\Sigma_t$ is $|\Sigma_t|^{-(p+1)}$, we obtain the density of $\Sigma_t$ as
$f(\Sigma_t|\Sigma_{t-1},A)=f(\Phi_t|\Phi_{t-1},A)|\Sigma_t|^{-(p+1)}$. Thus, from the above and by taking the logarithm in \eqref{logp1}, the log posterior function is
\begin{gather}
LP=3Np\log k -\frac{1}{2}\textrm{trace}(AF_0A'\Sigma_0^{-1})-\frac{2n+p}{2}\log |\Sigma_0|
-\frac{1}{2}\sum_{t=1}^N (y_t-\mu)'\Sigma_t^{-1}(y_t-\mu) \nonumber \\ -\frac{3p+2}{2}\sum_{t=1}^N\log |\Sigma_t| +(p+1)\sum_{t=1}^N \log |\Sigma_t^{-1}-\Lambda_t| - \frac{p}{2}\sum_{t=1}^N \log \xi_t, \label{logp2}
\end{gather}
where all constants are ignored, except $3Np\log k$. The reason we keep this constant is that $k$ depends on $\delta$.

The above log posterior is given conditionally on $A$. We can obtain a value of $LP$ if we replace $\Sigma_t$ and $A$ $(t=1,\ldots,N)$, by the estimates $\hat{\Sigma}_t$ and $\hat{A}$, where the former may be the mean or the mode of $\Sigma_t|D_t,A=\hat{A}$, both of which being routinely obtained by the posterior inverted Wishart densities. Then we can compare two models, which differ in the values of $\delta$'s, by using the principle of maximum log posterior. In the same lines of thinking, we may select the optimum $\delta$ that maximizes the above log posterior.

\subsection{Bayes factors}

Here we discuss Bayes factors and in particular we focus on sequential Bayes factors, which are introduced in West (1986) and discussed in detail in West and Harrison (1997, \S11.4). Bayes factors, as reviewed in Kass and Raftery (1995) and discussed in Gamerman and Lopes (2006, \S2.6) and in Robert (2007, \S7.2.2), are basically the posterior odd ratio of two models $\mathcal{M}_1$ and $\mathcal{M}_2$ (which are in competition) over the prior odd ratio. For sequential application, at each time $t$, the Bayes factor is defined by $BF_t=f(y_t|D_{t-1},\mathcal{M}_1)/f(y_t|D_{t-1},\mathcal{M}_2)$, see for example West (1986) for more details. Considering the above definition of $BF_t$, one has to compare it with 1 ($BF_t$ values larger than 1 indicate preference of $\mathcal{M}_1$, $BF_t$ values smaller than 1 indicate preference of $\mathcal{M}_2$ and $BF_t$ values equal to 1 indicate that the two models are equivalent, in the sense they both have the same predictive ability). One possibility for $\mathcal{M}_1,\mathcal{M}_2$ is to differ in their respective discount factors, $\delta_1,\delta_2$, in which case the Bayes factor at $t$ is
$$
BF_t=\frac{\Gamma((\delta_1n_1+p)/2) \Gamma(\delta_2n_2/2) |k_1\hat{A}_1F_{1,t-1}\hat{A}_1'+\Lambda_t|^{1/2} (1+e_{1t}'(k_1\hat{A}_1F_{1,t-1}\hat{A}_1'+\Lambda_t)e_{1t})^{-(\delta_1n_1+p)/2} }{
\Gamma((\delta_2n_2+p)/2) \Gamma(\delta_1n_1/2) |k_2\hat{A}_2F_{2,t-1}\hat{A}_2'+\Lambda_t|^{1/2} (1+e_{2t}'(k_2\hat{A}_2F_{2,t-1}\hat{A}_2'+\Lambda_t)e_{2t})^{-(\delta_2n_2+p)/2} },
$$
where $n_j,k_j,\hat{A}_j,F_{j,t-1}$ are the respective values of $n,k,\hat{A},F_{t-1}$, for $\delta=\delta_j$ and $j=1,2$. One may consider a monitoring procedure as those described in West (1986) and based on sequential application of $BF_t, t=1,2,\ldots$, or consider some rules on threshold values for the average Bayes factor $BF=N^{-1}\sum_{t=1}^N BF_t$, e.g. the rules of Jeffreys (1961), which are discussed in detail in Kass and Raftery (1995); see also Robert (2007, p. 228).

\subsection{Minimum time-averaged portfolio risk}

We consider the minimum time-averaged portfolio risk as a criterion, which selects the volatility estimator with smallest sampling variance. For this to end, we employ a sequential version of Markowitz (1959) mean-variance unconstrained optimization (using as loadings for the volatility the out of sample predictions at time $t$). Sequential portfolio selection aims to find at each time $t$ an
optimal weight vector $w_t$ to minimize the variance of the portfolio return $r_t=w_t'y_t$, i.e. minimize
$\var(r_t|D_{t-1})=w_t'\hat{\Sigma}_tw_t$, where $\hat{\Sigma}_t$ is the
one-step forecast covariance matrix of $y_t|D_{t-1}$. The unconstrained portfolio strategy
computes the optimal weights as
$$
w_t=\frac{m\hat{\Sigma}_t^{-1}\mu}{\mu'\hat{\Sigma}_t^{-1}\mu},
$$
where the expected return $w_t'\mu=m$ is assumed to be time-invariant. Considering no transaction costs, the realized return $r_t=w_t'y_t$
can be used to visually assess the performance of the allocation of the weights $w_t$
Similar portfolio allocation strategies, including constrained portfolio selection, are discussed in Aguilar and West (2000), Soyer and Tanyeri (2006), Han (2006) and in references therein. 

Adopting this criterion, with two variance estimators, producing portfolio variances $s_t^{(A)}=\var(r_t|D_{t-1},\textrm{estimator A})$ and $s_t^{(B)}=\var(r_t|D_{t-1},\textrm{estimator B})$, we would select estimator A, if $N^{-1}\sum_{t=1}^N s_t^{(A)}<N^{-1}\sum_{t=1}^N s_t^{(B)}$. Given a single model, we can apply the same principle to choose over discount factors $\delta$ or other model components.

\section{Simulation study}\label{simulations}

\begin{table}
\caption{Monte Carlo means with standard deviations (in brackets) for the Frobenius distance of estimated mode volatility and true value of the volatility, for 3 scenarios (scenario 1 for UWAR(1) of the precision volatility, scenario 2 for UWAR(2) of the precision volatility and scenario 3 for UWAR(1) of the volatility).}\label{table:sim}
\begin{center}
\begin{tabular}{l|c|c|c}
 & Scenario 1 & Scenario 2 & Scenario 3 \\ \hline $p=3$ & 0.0001 (0.001) & 0.0008 (0.002) & 0.0010 (0.001) \\
 $p=10$ & 0.0003 (0.001) & 0.0013 (0.003) & 0.0018 (0.002) \\  $p=30$ & 0.0007 (0.002) & 0.0025 (0.005) & 0.0033 (0.001)
\end{tabular}
\end{center}
\end{table}

In this section we carry out Monte Carlo experiments on 3 different simulated sequences $\{\Sigma_{it}\}$ $i=1,2,3$, in order to assess the efficiency of the proposed estimation approach, based on the UWAR(1) model. $\{\Sigma_{1t}^{-1}\}$ is generated from a UWAR(1) process, $\{\Sigma_{2t}^{-1}\}$ is generated from a UWAR(2) process, and $\{\Sigma_{3t}\}$ is generated from a UWAR(1) process. Under these three scenarios, we use the estimation proposed in section \ref{inference} using a UWAR(1) process for the precision of the volatility, and thus in scenarios 2 and 3 we use the ``wrong model'', while in scenario 1 we use the ``true model''. In each case matrix $A$ is randomly generated from a Gaussian matrix-variate distribution and a true value of $\delta=0.8$ has been used. We repeat the experiments for $p=3$, $p=10$ and $p=30$ (dimension of the covariance matrices) and we generate time series $\{y_t\}$ from model (\ref{model1}) with $\mu=0$. The time series length of each simulation is $N=1000$ and the Monte Carlo sample size is set to 100. Reported is the averaged (over all Monte Carlo samples and over time points $101\leq t\leq 1000$) Frobenius volatility distance (defined as the square root of the sum of the squared differences of the estimated volatility from the true simulated volatility). For the estimated volatility the out of sample approximate mode of the posterior distribution of the volatility is used. Reported also is the related Monte Carlo standard deviation. We have used the first 100 observations of each Monte Carlo sample to specify the prior $F_0$ (see also section \ref{empiricalresults} below, which discusses this prior setting for real data sets). For the estimation of $\Sigma_t$, the true value $\delta=0.8$ is used. $\delta$ is the most sensitive parameter here, for the specification of which the criteria discussed in section \ref{diagnostics} may be used (see also section \ref{empiricalresults}).  We note from Table \ref{table:sim} that the estimated averaged distances are small. As the dimension of the covariance matrix increases, the power of the estimation decreases, but still with reasonable results for $p=30$. Also, when the true model is assumed (first column of the table) the performance of the model is better compared to that from Scenarios 2 and 3. These results illustrate the performance of the proposed model, although more detailed consideration of simulation should be needed for a more conclusive result, e.g. in order to learn about the sensitivity of $\delta$. The modelling approach of this paper, allows the simulation and estimation of medium dimensional time-varying covariance matrices (either for the purpose of volatility estimation or more generally), which is a difficult task, as it is pointed out by many authors, see e.g. Gourieroux {\it et al.} (2009).

\section{Foreign exchange rates}\label{empiricalresults}

\subsection{The data}

In this section we present an analysis of five foreign exchange rates
\emph{vis-$\grave{a}$-vis} the US dollar. The exchange rates are the Canadian dollar (CAD),
Euro (EUR), Japanese Yen (JPY), British pound (GBP) and Australian dollar (AUD), all expressed as
number of units of the foreign currency per US dollar. The sample
period runs from $4$ January $1999$ until $31$ December $2009$, and
corresponds to $2760$ observations, sampled at daily frequencies.
This data set was obtained from the Pacific Exchange Rate Service of the University of
British Columbia ({\tt http://fx.sauder.ubc.ca/}).

To begin with, data is transformed to log returns. In the first two years (4 January 1999 to 31 December 2001) we use the data for pre-processing
purposes, in order to obtain sample estimates for $\mu$ and $\Sigma_0$. Then, starting at 2 January 2002 we run the volatility algorithm, in order to obtain forecasts of the volatility matrix.

\subsection{Description of competing models}

Here we consider four models, all adopting model specification (\ref{model1}) with (a) $\Sigma_t^{-1}$ following a UWAR(1) process (this model is referred to as UWAR), (b) $\Sigma_t^{-1}$ following Soyer and Tanyeri (2006) random walk model, being a UWAR(1) model with $A=I_p$, (this model is referred to as RW) (c) $\Sigma_t$ following the
Wishart specification of Philipov and Glickman (2006) (this is referred to as PGWAR) and (d) $\Sigma_t$ following the dynamic conditional correlation GARCH models of Engle (2002) (referred to as DCC).

The DCC specification (Engle, 2002) sets $\Sigma_t=D_tR_tD_t$, where $D_t$ is the diagonal matrix with elements $\sigma_{11,t}^{1/2},\ldots,\sigma_{pp,t}^{1/2}$ and $R_t$ is the dynamic correlation matrix, having as off-diagonal elements the correlations of $y_{it}$ and $y_{jt}$ and units as diagonal elements, where $\Sigma_t=(\sigma_{ij,t})$ and $y_t=(y_{1t},\ldots,y_{pt})'$. In other words the DCC specification combines time-varying variances (via $D_t$) and time-varying correlations (via $R_t$). For each of the squared diagonal elements of $D_t$ a GARCH(1,1) process is used and $R_t$ is modelled using exponentially smoothed standardized GARCH(1,1) residuals. Thus, under the DCC, the process of the elements of $\Sigma_t$ consists of autoregressive components of previous variances, correlations, and squared observed returns, while under the UWAR(1) specification, the process of the elements of $\Sigma_t$ consists of autoregressive components of previous variances and covariances (see equation \eqref{volprocess:sigma}). In the UWAR specification past volatility matrices as being stochastic, carry vital information via their conditional distribution, while in the DCC specification, such information is carried via explicit specification of their squared observed returns and of the latent structure of the unknown GARCH components. Another major difference, is that since the DCC inference is performed through likelihood-based estimation methods, the DCC is aimed at off-line estimation (when all data is available), while the UWAR can be applied and indeed in this paper it is targeted at on-line application.

Comparing different models that use Bayesian and non-Bayesian methods is a challenging task; some of the issues involved are reported in Dan\'{i}elsson (1998) who uses the likelihood function as a means of model comparison. In this paper (a) we compare two Bayesian models (UWAR and RW) using Bayes factors, the log-posterior function and the minimum time-averaged risk and (b) we use the Sharpe ratio and the minimum time averaged portfolio risk to compare models UWAR with PGWAR and UWAR with DCC.

\subsection{Empirical results}

Table \ref{table1} compares the performance of UWAR and RW models (using the log-posterior and the time average minimum portfolio risk) over a set of discount factors $\delta$ in the range $(0.7,1)$; for UWAR a vague Gaussian prior for $A$ is used with $M_A=0$, $V_A=W_A=1000I_5$, and $\Lambda_t=0$, for all $t$. We notice that the best performer is the UWAR with $\delta=0.7$, having largest log posterior function and minimum time averaged portfolio risk. The UWAR model with $\delta=0.7$ was also the best performer considering the Bayes factor of this model with $\delta=0.7$, vs the UWAR models with values of $\delta=0.75,0.8,0.85,0.9,0.95,0.98$ (average Bayes factor values 10.01, 15.9, 18.2, 23.5, 27.9, 33.02, respectively). The Bayes factor criterion also favoured UWAR model with $\delta=0.7$ when comparing it with any of the RW model, with any value of $\delta$ in the above range; the smallest of the average of the Bayes factor was 19.35. Consulting the above criteria (log-posterior function, time averaged portfolio risk and average Bayes factor), we conclude that UWAR outperforms RW and this basically illustrates the improvement we obtain by estimating $A$ (in the UWAR) as opposed to set it naively to $I_p$.

\begin{table}[ht]
\caption{Performance of UWAR and RW models for a set of discount factors $\delta$. Shown are the log posterior function (LP) and the time-averaged portfolio risk (Risk).}\label{table1}
\begin{tabular}{ll|rrrrrrr}
 & & & & & $\delta$ & & & \\ & & 0.7 & 0.75 & 0.8 & 0.85 & 0.9 & 0.95 & 0.98 \\ \hline UWAR & LP & 829259.2 & 805312.4 & 773072.8 & 726744.8 & 664228.6 & 503922 & 276256.9 \\ & Risk & 0.0013 & 0.0018 & 0.0019 & 0.0022 & 0.0028 & 0.0049 & 0.011 \\ \hline RW & LP & 817053.8 & 792433.3 & 759080.9 & 710748.5 & 632475.9 & 472273.9 & 211837.3 \\ & Risk & 0.0193 & 0.0209 & 0.0238 & 0.0286 & 0.0379 & 0.0678 & 0.1665 \\
\end{tabular}
\end{table}

As far as comparison with the other two models is concerned, firstly for the PGWAR we adopt
the efficient Gibbs sampler described in Philipov and Glickman (2006). The Gibbs sampler
burn-in stage is set to 1000 iterations. As in the above reference, at each time $t$, posterior
samples of 2000 draws are taken after the initial 1000 burn-in iterations. Finally, a Monte Carlo average of the mode of these samples is obtained and this is loaded onto the portfolio
exercise, yielding a time averaged portfolio risk 0.0012. This value is slightly smaller than that of
UWAR, however, the disadvantage of the PGWAR model is that it requires Gibbs sampling for
2008 time points, which is time consuming.

\begin{figure}[t]
\begin{center}
\includegraphics[scale=0.65]{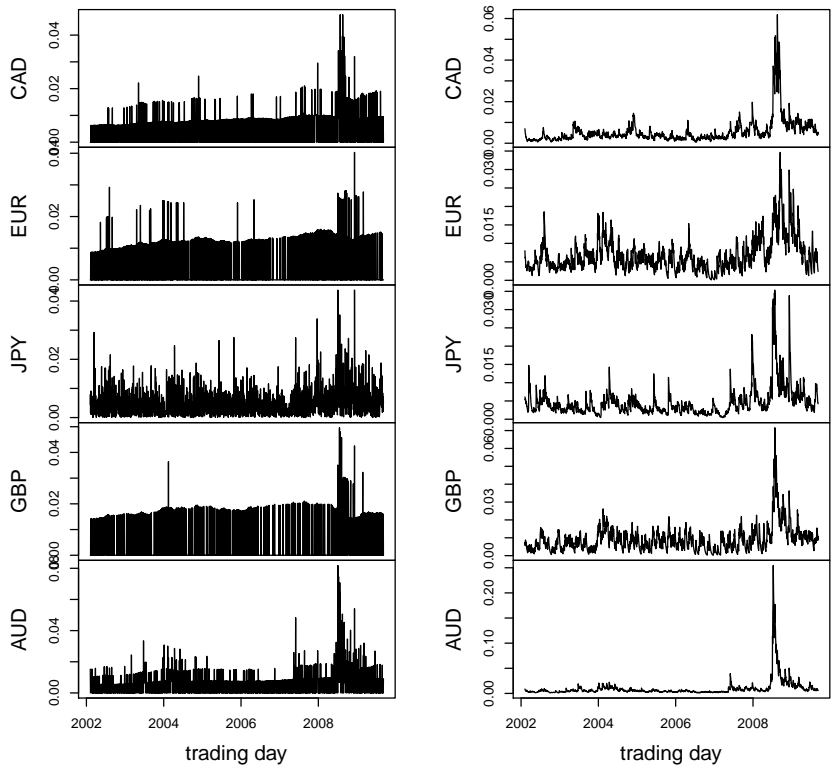}
\caption{Absolute returns and standard deviations of the out of sample predicted volatility, for the UWAR model with $\delta=0.7$.}
\label{fig1}
\end{center}
\end{figure}

\begin{figure}[t]
\begin{center}
\includegraphics[scale=0.65]{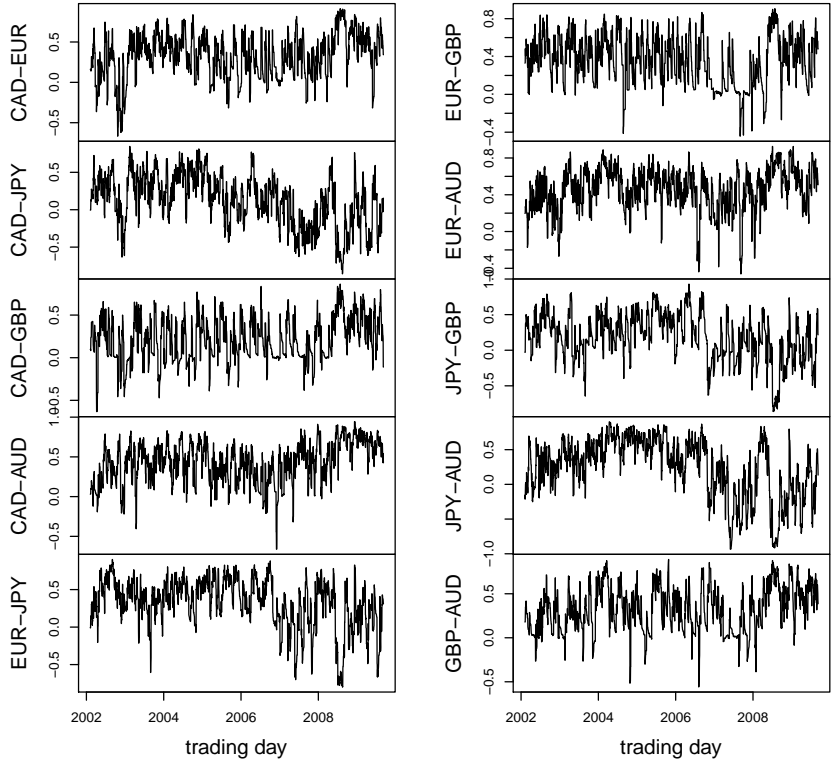}
\caption{Out of sample predictions of the cross-correlations between the five exchange rates, for the UWAR model with $\delta=0.7$.  }
\label{fig2}
\end{center}
\end{figure}

\begin{figure}[h]
\begin{center}
\includegraphics[scale=0.65]{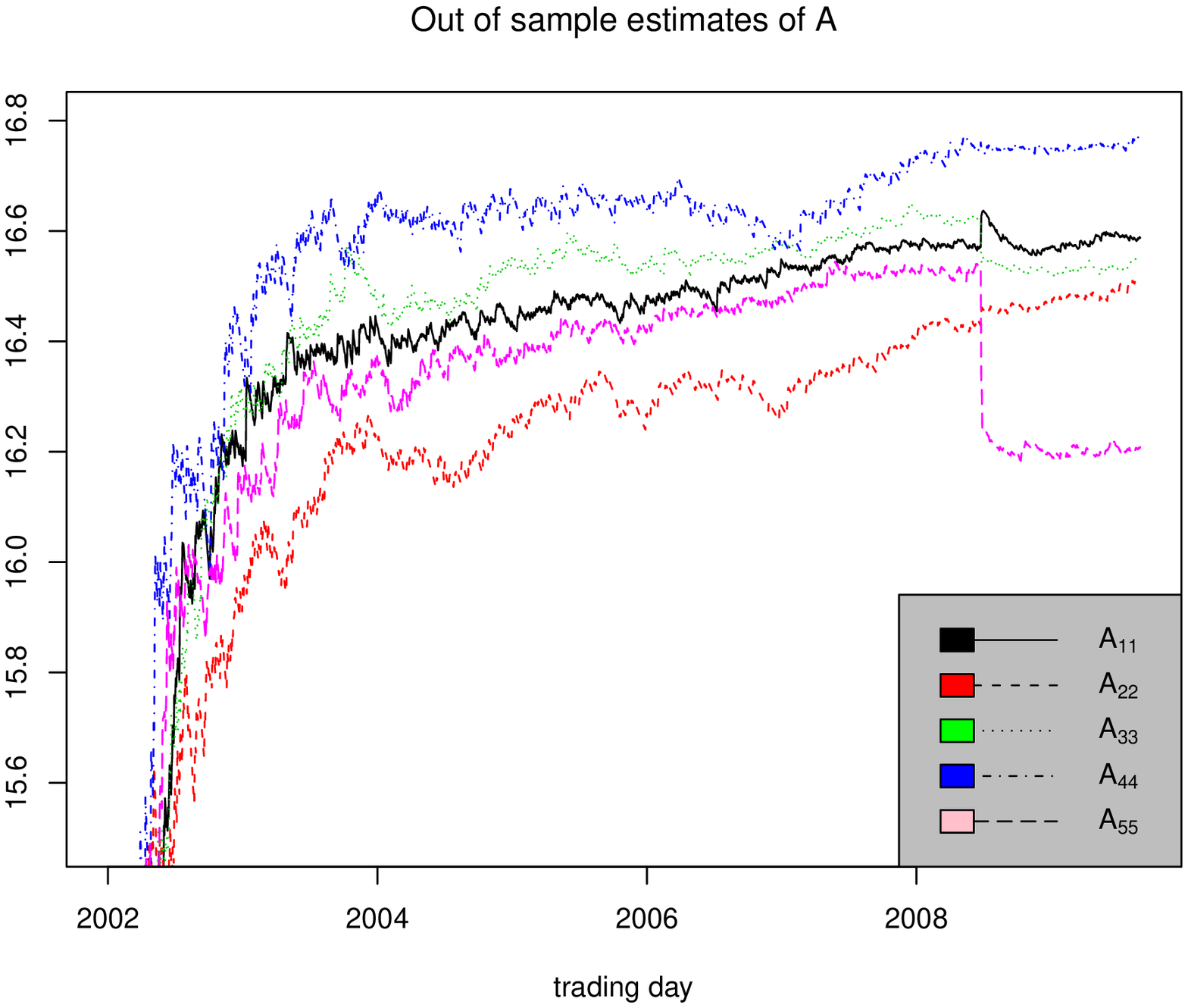}
\caption{Out of sample estimates of the diagonal elements $A_{ii}$ of $A=\{A_{ij}\}$, for the UWAR model with $\delta=0.7$. }
\label{fig3}
\end{center}
\end{figure}

A similar exercise was carried out regarding the DCC model with resulting averaged portfolio risk equal to 0.0019, which is larger than that of the UWAR. Comparing further the four models we find that the average conditional Sharpe ratio of the UWAR (with $\delta=0.7$), the RW (with $\delta=0.7$), the PGWAR and the DCC was 0.945, 0.566, 0.947 and 0.839, illustrating that the UWAR and the PGWAR are the best performers, using this criterion. We then conclude here that overall the UWAR is the best performer, although the PGWAR also puts a strong performance. For the UWAR model, Figure \ref{fig1} shows the absolute returns together with the out of sample predicted marginal volatilities (the diagonal elements of the predicted volatility matrix $\hat{\Sigma}_t$, conditioned upon information $D_{t-1}$ sequentially for $t=1,\ldots,N$ starting at 2 January 2002) and Figure \ref{fig2} shows the out of sample predicted correlations. Figure \ref{fig1} indicates the good out of sample forecasting performance of the volatility, while Figure \ref{fig2} shows the dynamics of the correlation. Figure \ref{fig3} shows the estimates of the diagonal elements of $A=(A_{ij})_{i,j=1,\ldots,5}$. We note that $A_{11}$ and $A_{55}$ indicate a structural change after 2008, which highlights the abrupt increase in the volatility at that period, being evident by the left panel of Figure \ref{fig1} for CAD (relevant to $A_{11}$) and AUD (relevant to $A_{55}$). We also note that initially, the values of $A_{ii}$ are centered around one ($A$ is the autocorrelation of the precision process $\{\Phi_t$\}). In Figure \ref{fig3}, we see that the $A_{ii}$'s gradually increase (the autocorrelations of the volatility process are $(A')^{-1}$ multiplied by a constant, see e.g. equation \eqref{volprocess:sigma}). Thus, after 2003 the estimated values of $A_{ii}$ are centered around 16.4, although for more conclusive comments one needs to look at the off-diagonal elements of $A$ too.  For the Newton-Raphson algorithm we have used a stoppage tolerance $Tol=0.0001$ and this was achieved for a minimum of 4 iterations and a maximum of 10 iterations.

\section{Concluding remarks}\label{conclusions}

This paper develops a new methodology for multivariate volatility estimation. Assuming the volatility matrix to be positive definite, the core of the methodology commences by considering that the stochastic evolution of the precision of the volatility follows a Wishart autoregressive process. The paper proposes inference conditional and unconditional on the autoregressive parameters. The proposed methodology does not rely on simulation-based methods (such as MCMC and particle filters) or on maximum likelihood estimation (such as the several GARCH procedures reported in Bauwens {\it et al.}, 2006), but still retains desirable complexity describing the dynamics of the volatility. This proposes an efficient, but realistic probabilistic setting, with application to medium dimensional financial data and to systems that real-time estimation is required. Recently, such systems have been much of the discussion, in the finance industry, such as in hedge funds and in other proprietary financial boutiques in which automatic or algorithmic trading is in high demand.

\section*{Acknowledgements}

I am grateful to two anonymous referees for their helpful comments, which led to a considerably improved version of the paper.

\renewcommand{\theequation}{A-\arabic{equation}} 
\setcounter{equation}{0}  

\section*{Appendix A: Singular multivariate beta distribution}

In this section we provide some details about the multivariate beta distribution mentioned in section \ref{inference:onA}. The Wishart and multivariate beta convolution is well known in the literature (a good account is given in Muirhead, 1982, Theorem 3.3.1), but Uhlig (1994) in his introduction demonstrates that for Wishart processes aimed at financial application, the aforementioned convolution is not suitable. Uhlig proposes the development of singular multivariate beta distribution, as a modelling mechanism to define random walk type stochastic process for Wishart matrices, retaining the desirable conjugacy between the Wishart and the beta distributions. Formally, the $p\times p$ matrix $B$ follows the singular beta distribution, if $B=(\U(X+Y)')^{-1}Y\U(X+Y)$, where $X\sim W_p(a,I_p)$, $Y\sim W_p(b,I_p)$, $X,Y$ are independent, and $\U(X+Y)$ denotes the upper triangular factor of the Choleski decomposition of $X+Y$, i.e. $X+Y=\U(X+Y)'\U(X+Y)$. In this definition, it is assumed that $a>p-1$ so that $X$ follows a non-singular Wishart distribution and the positive integer $b$ satisfies $1\leq b\leq p-1$, so that $Y$ follows a singular Wishart distribution. A similar argument can be made if $a\leq p-1$ is integer and $b>p-1$. In terms of notation we write $B\sim B_p(a/2,b/2)$ and the density of $B$, which is defined in the Steifel manifod, is
$$
f(B)=\frac{\pi^{-(pb+b^2)/2}\Gamma_p((a+1)/2)}{\Gamma_b(b/2)\Gamma_p(a/2)}|B|^{(a-p-1)/2} |L|^{(b-p-1)/2},
$$
where $L$ is the diagonal matrix with elements the positive eigenvalues of $I_p-B$, which are exactly $b$. If $b>p-1$, the density reduces to the non-singular multivariate beta density (Muirhead, 1982), in which case $|L|=|I_p-B|$.

The key property of the above distribution, is that if $\Phi\sim W_p(a+b,F)$ with $a>p-1$ and for some integer $b>0$, and if $B\sim B_p(a/2,b/2)$ independently of $\Phi$, then $\Phi^*=\U(\Phi)'B\U(\Phi)\sim W_p(a,F)$. This extends the Wishart and beta convolution, to allow situations where $a+b\leq 2p-2$, $a>p-1$ and $b$ a positive integer. The singular beta distribution has attracted considerable interest over the recent years, for further details of which the reader is referred to D\'{i}az-Garc\'{i}a and Guti\'{e}rrez (2008).

\section*{Appendix B: Proof of equation (\ref{deriv})}

Let $x_{ij}$ be the $(i,j)$th element of $X$ and write $D=BG+I_p$. It is
\begin{eqnarray*}
\frac{\partial \log |BXCX'+D|}{\partial x_{ij}} &=& \textrm{trace}\left((BXCX'+D)^{-1}\frac{\partial (BXCX'+D)}{\partial x_{ij}}\right) \\ &=&
\textrm{trace}((BXCX'+D)^{-1}Bu_iu_j'CX') + \textrm{trace}((BXCX'+D)^{-1}BXCu_ju_i') \\ &=& \textrm{trace}(CX'(BXCX'+D)^{-1}Bu_iu_j')+ \textrm{trace}
((BXCX'+D)^{-1}BXCu_ju_i') \\ &=& u_j'CX'(BXCX'+D)^{-1}Bu_i+u_i'(BXCX'+D)^{-1}BXCu_j,
\end{eqnarray*}
where $u_i=(0,\ldots,0,1,0,\ldots,0)'$, for $i=1,\ldots,p$, so that $x_{ij}=u_i'Xu_j$. Putting the above equation in matrix form we obtain
\begin{eqnarray*}
\frac{\partial \log |BXCX'+D|}{\partial X}&=&(CX'(BXCX'+D)^{-1}B)'+(BXCX'+D)^{-1}BXC\\ &=&(B(XCX'B+D')^{-1}+(BXCX'+D)^{-1}B)XC
\end{eqnarray*}
and the result follows by observing that matrix $(BXCX'+D)^{-1}B$ is symmetric, i.e.
\begin{eqnarray*}
B&=&B(XCX'B+GB+I_p)(XCX'B+GB+I_p)^{-1} \\ \Leftrightarrow B &=& (BXCX'+D)B(XCX'B+D')^{-1} \\ \Leftrightarrow
(BXCX'+D)^{-1}B &=& B(XCX'B+D')^{-1}.
\end{eqnarray*}

\end{document}